\documentclass{article}
\usepackage{spconf,amsmath,graphicx,hyperref}
\usepackage{cite}
\usepackage{amsmath,amssymb,amsfonts,latexsym}
\usepackage{amsthm}
\usepackage{braket}
\usepackage{graphicx}
\usepackage{textcomp}
\usepackage{xcolor}
\usepackage{mathtools}
\usepackage{multirow}
\usepackage{url}
\usepackage{booktabs}
\usepackage{hyperref}
\usepackage{lipsum}
\usepackage{nicematrix}
\usepackage[all]{xy}
\usepackage{tikz-cd}
\usepackage{tikz}
\usepackage{makecell}
\usepackage{caption}
\usepackage{subcaption}
\usepackage{quantikz}

\usetikzlibrary{calc}
\usetikzlibrary{intersections,shapes.arrows}
\usetikzlibrary{arrows.meta, positioning, backgrounds}

\usepgfmodule{nonlineartransformations}

\def \s{\sigma}

\def \mcal{\mathcal}

\title{Quantum Super-resolution by Adaptive Non-local Observables
}

\name{
  Hsin-Yi Lin$^{1}$ \quad
  Huan-Hsin Tseng$^{2}$\thanks{Code available at \url{https://github.com/hsinyilin19/ANO_SR}. This work is supported by Laboratory Directed Research and Development Program \#24-061 of Brookhaven National Laboratory.} \quad
  Samuel Yen-Chi Chen$^{3}$\thanks{The views expressed in this article are those of the authors and do not represent the views of Wells Fargo. This article is for informational purposes only. Nothing contained in this article should be construed as investment advice. Wells Fargo makes no express or implied warranties and expressly disclaims all legal, tax, and accounting implications related to this article.} \quad
  Shinjae Yoo$^{2}$
}
\address{
  $^{1}$ Seton Hall University, Department of Mathematics \& Computer Science, South Orange, NJ, USA \\
  $^{2}$ Brookhaven National Laboratory, AI \& ML Department, Upton, NY, USA \\
  $^{3}$ Wells Fargo, New York, NY, USA \\
  hsinyi.lin@shu.edu, htseng@bnl.gov, yen-chi.chen@wellsfargo.com, syjoo@bnl.gov
}

\begin{document}
\maketitle

\begin{abstract}

Super-resolution (SR) seeks to reconstruct high-resolution (HR) data from low-resolution (LR) observations. Classical deep learning methods have advanced SR substantially, but require increasingly deeper networks, large datasets, and heavy computation to capture fine-grained correlations. In this work, we present the \emph{first study} to investigate quantum circuits for SR. We propose a framework based on Variational Quantum Circuits (VQCs) with \emph{Adaptive Non-Local Observable} (ANO) measurements. Unlike conventional VQCs with fixed Pauli readouts, ANO introduces trainable multi-qubit Hermitian observables, allowing the measurement process to adapt during training. This design leverages the high-dimensional Hilbert space of quantum systems and the representational structure provided by entanglement and superposition. Experiments demonstrate that ANO-VQCs achieve up to five-fold higher resolution with a relatively small model size, suggesting a promising new direction at the intersection of quantum machine learning and super-resolution.
\end{abstract}

\section{\label{sec:Indroduction}Introduction}

Image super-resolution (SR) aims to reconstruct a high-resolution (HR) image from one or more low-resolution (LR) counterparts. Applications of SR span diverse fields, from natural photography \cite{dong2015srcnn}, satellite and remote sensing \cite{liu2023ediffsr}, to medical imaging \cite{chung2022mri}, where fine structural details are critical.  

Classical approaches to SR include interpolation-based and sparse-representation methods \cite{yang2010image, yang2014benchmark}, but deep learning has driven dramatic improvements in recent years. Early convolutional neural network (CNN) models such as SRCNN \cite{dong2015srcnn}, FSRCNN \cite{dong2016fsrcnn}, and VDSR \cite{kim2016vdsr} demonstrated significant performance gains. Subsequent architectures incorporated residual connections \cite{ledig2017srgan, zhang2018rcan}, dense connectivity \cite{tong2017dscn}, attention mechanisms \cite{niu2020han}, and transformer-based designs such as SwinIR \cite{liang2021swinir}. Beyond regression models, generative frameworks have become dominant: GAN-based SR methods \cite{ledig2017srgan, wang2018esrgan} and more recently diffusion models \cite{saharia2022sr3} achieve perceptually realistic HR reconstructions, albeit with high computational cost and extensive data requirements.  

Quantum computing offers a fundamentally different paradigm for information processing, with the potential to complement or transcend classical architectures. In particular, Variational Quantum Circuits (VQCs), which are hybrid quantum-classical models optimized through variational principles, have shown promise in quantum machine learning \cite{cerezo2021vqa, mcclean2016theory, bharti2022nisq}. Conventional VQCs, however, rely on fixed local observables (often Pauli operators) for measurement, which limits expressivity and restricts access to the exponentially rich Hilbert space.  

Recently, an \emph{Adaptive Non-Local Observable} (ANO) framework was introduced \cite{lin2025ano}, inspired by the Heisenberg picture, where the measurement operators themselves are treated as trainable Hermitian observables acting on multiple qubits. By learning both circuit parameters and observables, ANO-VQCs expand the representational capacity of quantum neural networks, enabling richer qubit interactions, enhanced feature mixing, and parameter efficiency. Importantly, non-local observables allow the readout to span different equivalence classes of Hermitian operators, thus capturing a broader range of eigenvalue spectra and providing more informative measurements than fixed Pauli observables.  

In this work, we propose applying ANO-VQCs to image super-resolution. Our central insight is that non-local observables naturally serve as higher-resolution ``lenses'' for quantum systems: by enabling measurements across entangled multi-qubit subspaces, ANO-VQCs can extract fine-grained information that corresponds to HR details. This approach leverages the intrinsic high dimensionality of quantum Hilbert space to realize perceptual improvements in SR without requiring prohibitively deep circuits or large qubit counts. By bridging recent advances in SR research \cite{wang2021survey} with adaptive quantum measurement strategies \cite{lin2025ano}, this work positions ANO-based quantum circuits as a novel and resource-efficient paradigm for super-resolution.

% ============================ Section ============================ %
\section{Adaptive Non-local Observable VQC (ANO-VQC)}\label{sec_method}

\subsection{General VQC Structures}\label{subsec_VQC structure}

The VQC is an approach to realize QML by seeking proper unitary transformations in qubit systems to fit and learn data. 

Let $\mathcal{D} = \{(x^{(j)}, y^{(j)}) \, | \, x^{(j)} \in \mathbb{R}^n, y^{(j)} \in \mathbb{R} \}$ be a set of classical data with $x^{(j)}$ as an input of sample $j$ and $y^{(j)}$ as the corresponding ground truth. 

The VQC fits $\mathcal{D}$ by three steps: \textit{encoding, variation, and measurement} (Fig.~\ref{fig: Variational circuits}). Denote the set of unitary transformations in Hilbert space $\mathcal{H}^n$ by $\mathcal{U}(\mathcal{H}^n)$. The encoding step maps a classical input $x \in \mathbb{R}^n$ into a unitary $V(x) \in \mathcal{U}(\mathcal{H}^n)$. When acting on a random initial state $\ket{\psi_0} \in \mathcal{H}^n$, $\ket{\psi_x} := V(x)\ket{\psi_0}$ is called the \textbf{encoded state} to carry the information of $x$. The variation step applies a parameterized unitary $U(\theta)$ to the encoded state $\ket{\psi_x}$, yielding $\ket{\psi_{\theta,x}} := U(\theta)\ket{\psi_x}$. The tunable parameters $\theta$ play a role analogous to weights in classical neural networks. By varying $\theta$, the resulting state $\ket{\psi_{\theta,x}}$ traverses different regions of the Hilbert space $\mathcal{H}^n$ in search of a representation that best captures the structure of the learning task.

Final measurement chooses a Hermitian matrix $H^{\dagger} = H$ to compute the following bilinear form as the model prediction,
\begin{equation}\label{E: VQC output}
f_\theta(x) := \bra{\psi_{\theta,x}} H \ket{\psi_{\theta,x}} = \bra{\psi_0} V^{\dagger}(x) \underbracket{U^{\dagger}(\theta) H U(\theta)} V(x) \ket{\psi_0}
\end{equation}
to be compared with the target value $y$ via a loss function,
\begin{equation}\label{E: loss}
L(\theta; \mathcal{D}) = \frac{1}{| \mcal{D} |} \sum_{j=1}^{| \mcal{D} |} \| f_\theta(x^{(j)}) - y^{(j)} \| ^2
\end{equation}
In Quantum Mechanics, a Hermitian matrix $H$ for measurement is also called an \textbf{observable}. In the VQC, an observable is chosen to be \textit{fixed}, typically from Pauli matrices $\mathcal{P} = \{I, \s_1, \s_2, \s_3 \}$. 

On the other hand, the parametrized $U(\theta)$ is the \textit{only} learnable component to fit $\mcal{D}$ by minimizing loss \eqref{E: loss}, while the encoding map $V$ and the measurement Hermitian $H$ remain fixed. General quantum circuit compositions employ Hadamard gate $\texttt{H}$, CNOT gate $\mcal{C}$, and rotation gates $ \{e^{ -\frac{i}{2} \phi \s_1}, e^{ -\frac{i}{2} \phi \s_2}, e^{ -\frac{i}{2} \phi \s_3}\}$ generated by Pauli $\mathcal{P}$ with angle $\phi$. For example, a choice of an encoding matrix can be,
\begin{equation}\label{E: encoding V}
    V(x) = \bigotimes_{q=1}^n \left(e^{ -\frac{i}{2} x_q \, \s_{k_q} } \circ \texttt{H}\right)
\end{equation}
and a variational circuit
\begin{equation}\label{E: variational U}
    U(\theta) = \prod_{\ell=1}^L  \left( \bigotimes_{q=1}^n e^{ -\frac{i}{2} \theta_q^{(\ell)} \s_q  } \right) \circ  \mathcal{C}_{\ell} \in \mcal{U}(\mcal{H}^n)
\end{equation}
with tunable parameters $\theta = (\theta_1^{(1)} , \ldots, \theta_n^{(L)}) \in \mathbb{R}^{n \times L}$ (Fig.~\ref{fig: Variational circuits}).

% -------------- Figure -------------- %
\begin{figure}[htbp]
 % \vskip -0.1in
  \centering
  \scalebox{0.9}{
    \begin{quantikz}[row sep=0.1cm, column sep=0.6cm]
        % Qubits 1 and 2 form the first CNOT pair; qubits 3 and 4 form the second.
        \lstick{$\ket{0}$} &\gate{H} \gategroup[wires=4,steps=2,style={dashed,rounded corners,fill=blue!20,inner xsep=2pt},background]{Encoding $V$} & \gate{R_y(x_1)} & \ctrl{1} \gategroup[wires=4,steps=3,style={dashed,rounded corners,fill=magenta!20,inner xsep=2pt},background]{(Variational $U$) $\times L$} & \qw &\gate{R_y(\theta_1)}  & \meter{} \\
        \lstick{$\ket{0}$} &\gate{H} & \gate{R_y(x_2)} & \targ{}  & \ctrl{1} &\gate{R_y(\theta_2)}  & \meter{} \\
        \lstick{$\ket{0}$} &\gate{H} & \gate{R_y(x_3)} & \ctrl{1} & \targ{}  &\gate{R_y(\theta_3)}  & \meter{} \\
        \lstick{$\ket{0}$} &\gate{H}  & \gate{R_y(x_4)} & \targ{}  & \qw      &\gate{R_y(\theta_4)}  & \meter{}
    \end{quantikz}
    }
  \caption{A VQC diagrm of (\ref{E: encoding V}), (\ref{E: variational U}), which is also implemented in Sec.~\ref{sec_exp_results}. The variational block represented by a pink box is repeated $L$ times to increase the circuit depth.}
  \label{fig: Variational circuits}
   \vskip -0.1in
\end{figure}
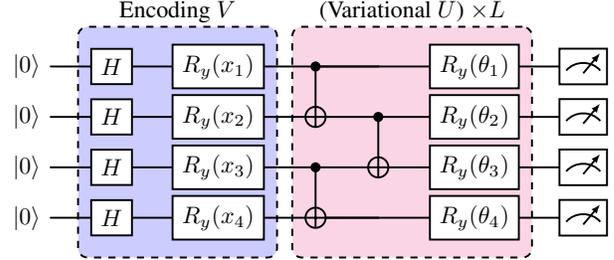
% -------------- Figure -------------- %

% -------------- Figure -------------- %
\begin{figure}[htbp]
    \centering 
\begin{tikzpicture}[scale=0.8]

  \def\Mcurve{
    plot [smooth cycle] coordinates {(0,0) (3,0.3) (3.5,1.5) (3,2.7) (0,3) (-1,1.5)}
  }
  
  \begin{scope}
    \clip \Mcurve;
    \fill[blue!20, opacity=0.6] (-2,0) rectangle (5,1);
  \end{scope}
  % Middle partition: y in [1,2]
  \begin{scope}
    \clip \Mcurve;
    \fill[green!20, opacity=0.6] (-2,1) rectangle (5,2);
  \end{scope}
  % Top partition: y in [2,3]
  \begin{scope}
    \clip \Mcurve;
    \fill[red!20, opacity=0.6] (-2,2) rectangle (5,3);
  \end{scope}

  % Draw horizontal dashed boundaries (at y=1 and y=2) clipped to the manifold.
  \begin{scope}
    \clip \Mcurve;
    \draw[dashed, thick] (-2,1) -- (5,1);
    \draw[dashed, thick] (-2,2) -- (5,2);
  \end{scope}

  % Draw the boundary of the curved manifold M.
  \draw[thick] \Mcurve;
  \node at (2.2,3.2) {$\mathbb{H}(n)$};

  % Label each equivalence class (partition) inside M.
  \node at (1.3,2.5) {$H_0$\footnotesize{ (VQC)}};
  \node at (0.8,1.5) {$H_1$};
  \node at (0.8,0.5) {$H_2$};
\end{tikzpicture}
\caption{Let $\mathbb{H}(n)$ denote the space of all observables represented by ANO. The conventional VQC with fixed Pauli measurements becomes a special case (an equivalent subclass) within the ANO function classes; see~\cite{lin2025ano} for details.}
\label{fig: Hermitian equivalent classes}
\end{figure}
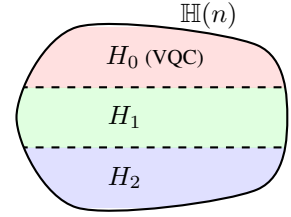
% -------------- Figure -------------- %

It is demonstrated in \cite{chen2025learning} that relaxing the constraint of fixed Pauli measurements with learnable observables leads to a marked enhancement in model performance. 

It is further analyzed in \cite{lin2025ano} that through the adaptation of non-local observables, the VQC formalism becomes a special case of the ANO function class (Fig.~\ref{fig: Hermitian equivalent classes}). Thus, $k$-local ANO of the following form unifies and generalizes the standard VQC,
\begin{equation}\label{E: non-local Hermitian}
  H(\phi) =   \begin{pmatrix}
c_{11} & a_{12} + i b_{12} & a_{13} + i b_{13} & \cdots & a_{1K} + i b_{1K}  \\
* & c_{22}  & a_{23} + i b_{23}  & \cdots & a_{2K} + i b_{2K}  \\
* & * & c_{33}  & \cdots & a_{3K} + i b_{3K}  \\
\vdots & \vdots & \vdots & \ddots & \vdots \\
* & * & * & \cdots & c_{KK}
\end{pmatrix}
\end{equation}
where $\phi = \left( a_{ij}, b_{ij}, c_{ii} \right)_{i, j=1}^K$ are arbitrary $K^2$ real parameters to be varied with $K = 2^k$, and the lower triangle entries are determined by the complex conjugates of the upper triangle such that $H(\phi) = (h_{ij}) = (\overline{h_{ji}}) = H^{\dagger}(\phi)$. (\ref{E: non-local Hermitian}) is a form of \textbf{$k$-local} observable in an $n$-qubit system ($k\leq n$).

The structural flexibility inherited from ANO permits the execution of complex tasks, such as resolution enhancement in image processing.

\begin{figure*}[htb]
\vskip -0.3in
\centering
\includegraphics[width=0.95\textwidth]{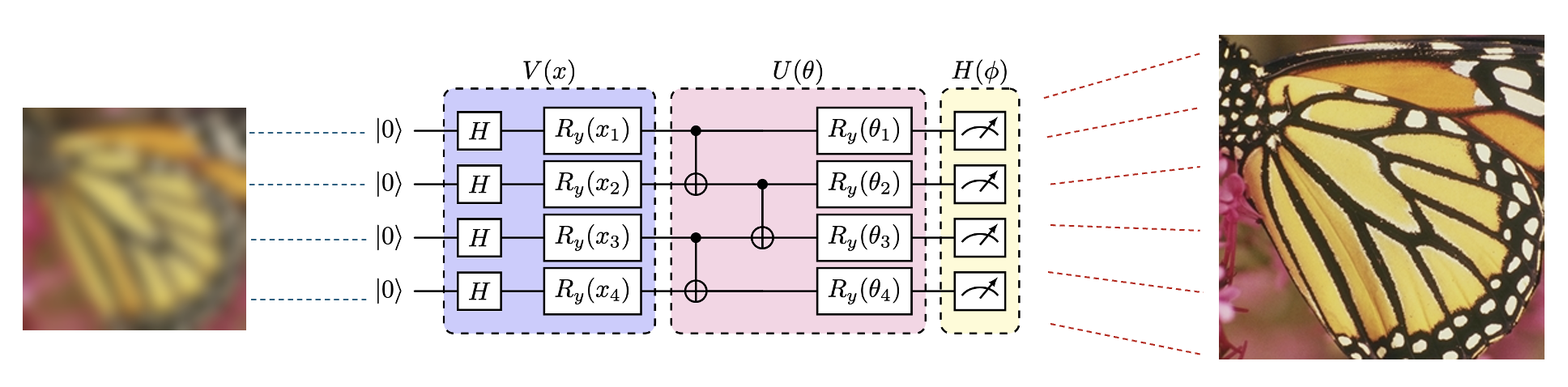}
\caption{\textbf{ANO-VQC for SR.} An LR input $x$ is encoded by $V(x)$ into state $V(x)\ket{\psi_0}$, transformed by variational layers $U(\theta)$, and finally measured by an adaptive $k$-local observable $H(\phi)$ to reconstruct the HR output.}
\label{fig:ano_vqc_sr_tworow}

\end{figure*}

% --------------------- Sub-Section --------------------- %
\subsection{ANO-VQC for SR}\label{subsec_ANO_VQC_SR}

Building upon the general VQC framework in Sec.~\ref{subsec_VQC structure}, we apply ANO-VQC architecture for the quantum realization of SR. The dimensionality of LR inputs is enlarged by harnessing the expressive power of variational transformations together with learnable non-local observables. This allows the network to reconstruct HR outputs from quantum measurements.

\subsubsection{\textbf{Workflow Structure}}  
The workflow of the ANO-VQC for SR is composed of the following stages:
\begin{enumerate}
    \item \textbf{Encoding of LR images:} Each LR input image is first vectorized and embedded into the $n$-qubit Hilbert space by an encoding unitary $V(x)$ as in \eqref{E: encoding V}. 
    \item \textbf{Variational transformation:} The encoded state is processed by variational unitaries $U(\theta)$ defined in \eqref{E: variational U}. The tunable parameters $\theta$ act to explore the quantum state space, preparing representations that are more suitable for the next measurement step. 
    \item \textbf{HR images by ANO:} We employ adaptive $k$-local observables $H(\phi)$ as in \eqref{E: non-local Hermitian}. Multi-qubit measurements are repeatedly performed, generating the predicted HR pixel values. 
\end{enumerate}

\subsubsection{\textbf{Learning Mechanism}} 
The trainable parameters are:  
\begin{itemize}
    \item \emph{variational rotation angles} $\theta=\left\{\theta^i\right\}$,
    \item \emph{adaptive Hermitian observable} $\phi=\left\{\phi^i\right\}$.  
\end{itemize}
The joint optimization of the variational parameters $\theta$ and the adaptive observable parameters $\phi$ is achieved through minimizing a reconstruction loss that measures the difference between the predicted outputs and the ground-truth HR images. To balance pixel-level fidelity with perceptual quality, we adopt a \textit{combined loss function}:  
\begin{equation}\label{eq:combined_loss}
\mathcal{L}(\theta, \phi) = c_1 \, \text{MSE} + c_2 \, \text{LPIPS},
\end{equation}
where $c_1, c_2 > 0$ are weighting coefficients. The mean squared error (MSE) term penalizes deviations in pixel intensities, ensuring accurate reconstruction of fine details. The Learned Perceptual Image Patch Similarity (LPIPS) term~\cite{zhang2018unreasonable} evaluates \textit{perceptual similarity} by comparing deep feature representations extracted from pretrained neural networks, thus aligning the reconstructed outputs with human visual perception.  

This combination allows the model to capture both \textbf{low-level accuracy} (via MSE) and \textbf{high-level perceptual quality} (via LPIPS), providing a more comprehensive objective for training ANO--VQC-based SR.

% ============================ Section ============================ %
\section{Experiments}\label{sec_exp_results}

% --------------------- Sub-Section --------------------- %
\subsection{Experimental setup}

Experiments were conducted on the \textbf{MNIST} dataset, consisting of $28\times28$ grayscale images of handwritten digits (0--9). 
Images were downsampled to $4\times4$ as LR inputs and upsampled to $12\times12$, $16\times16$, and $20\times20$ for $\times3$, $\times4$, and $\times5$ HR targets, respectively. 
The proposed ANO--VQC models were evaluated on these SR tasks, with both 2-local and 3-local adaptive observables implemented to assess the impact of non-local measurement depth on reconstruction fidelity and perceptual quality.

% --------------------- Sub-Section --------------------- %

\begin{table}[t]
    \centering
    \caption{\textbf{Super-resolution test performance metrics for 2-local and 3-local methods} Performance across different scaling factors showing degradation as scale increases.}
    \resizebox{0.9\columnwidth}{!}{%
    \begin{tabular}{p{0.12\linewidth} r | r r r r}
        \toprule
        \textbf{} & \textbf{Scale} & \textbf{MSE} $\downarrow$ & \textbf{LPIPS} $\downarrow$ & \textbf{PSNR} $\uparrow$ & \textbf{SSIM} $\uparrow$ \\
        \cmidrule(r){1-6}
        \multirow{3}{*}{2-local} & x3 & 0.42 & 0.16 & 24.13 & 0.84 \\
         & x4 & 0.62 & 0.19 & 22.32 & 0.76 \\
         & x5 & 0.80 & 0.22 & 21.18 & 0.66 \\
        \bottomrule
    \end{tabular}}
\label{tab:2local} 
\end{table}

\begin{table}[t]
    \centering
    \resizebox{0.9\columnwidth}{!}{%
    \begin{tabular}{p{0.12\linewidth} r | r r r r}
        \toprule
        \textbf{} & \textbf{Scale} & \textbf{MSE} $\downarrow$ & \textbf{LPIPS} $\downarrow$ & \textbf{PSNR} $\uparrow$ & \textbf{SSIM} $\uparrow$ \\
        \cmidrule(r){1-6}
        \multirow{3}{*}{3-local} & x3 & 0.35 & 0.17 & 24.85 & 0.87 \\
         & x4 & 0.53 & 0.20 & 23.04 & 0.80 \\
         & x5 & 0.69 & 0.23 & 21.83 & 0.70 \\
        \bottomrule
    \end{tabular}}
    \label{tab:3local}
\end{table}

\subsection{Results}

% -------------- Large Figure -------------- %
\begin{figure*}[htb]
  \centering
\includegraphics[width=0.8\textwidth]{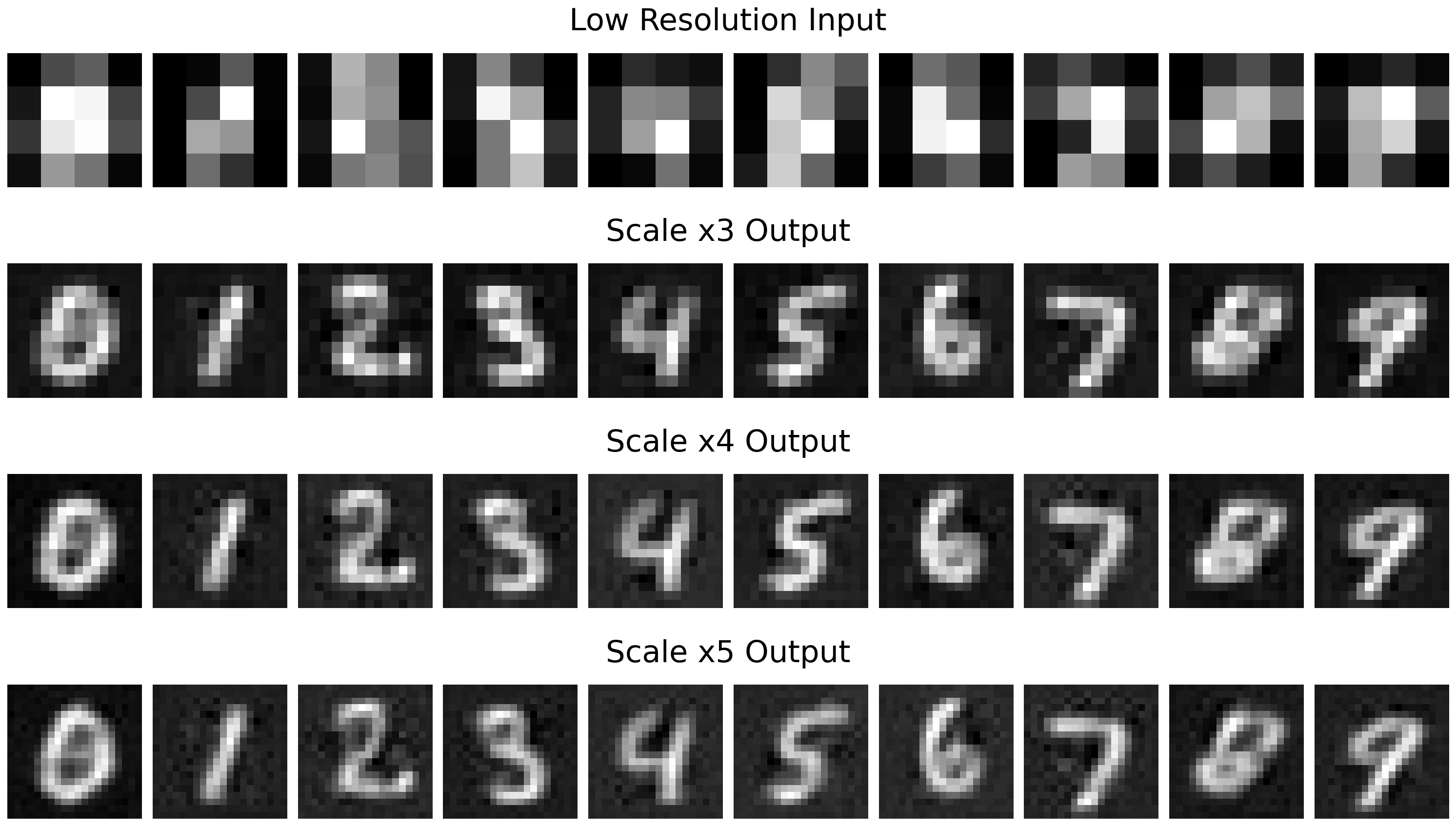}
  \caption{Super-resolution with 3-local ANO–VQC: $4 \times 4$ LR digit inputs (top) are reconstructed into $12 \times 12$, $16 \times 16$, and $20 \times 20$ HR outputs in the second, third, and fourth rows.}
  \label{fig:sr_results}
  \vskip -0.1in
\end{figure*}
% -------------- Large Figure -------------- %

Tables~\ref{tab:2local} summarize the quantitative metrics, including MSE, peak signal-to-noise ratio (PSNR), structural similarity (SSIM), and LPIPS. 
Across all scaling factors, the 3-local ANO--VQC consistently achieved lower MSE and higher PSNR and SSIM than the 2-local counterpart, indicating more accurate pixel-level reconstruction. 
However, its LPIPS values are slightly higher, suggesting a modest perceptual degradation compared to the 2-local case. 
This trade-off implies that deeper non-local observables enhance reconstruction fidelity but may slightly overfit high-frequency components, resulting in perceptually sharper yet less natural details.

The improvement is most evident for the $\times3$ SR task, where the 3-local model reached an MSE of $\textbf{0.35}$ and an SSIM of $\textbf{0.87}$, compared to $0.42$ and $0.84$ for the 2-local variant. 
As the scaling factor increases to $\times5$, both models show gradual degradation in reconstruction quality, a natural effect of larger upsampling ratios. Fig.~\ref{fig:sr_results} visualizes the SR outputs from the 3-local ANO-VQC.

Overall, the results highlight the advantage of using non-local measurement operators $H(\phi)$. By allowing multi-qubit Hermitian observables to adapt during training, the model explores a richer subspace of the Hilbert space, effectively enlarging the representational dimensionality.

The experimental results confirm that the proposed method effectively expands the expressive power of variational quantum circuits without requiring deeper layers or additional qubits. 
By optimizing both variational angles $\theta$ and Hermitian parameters $\phi$, the ANO-VQC jointly learns how to transform and how to observe the quantum state for faithful image reconstruction.

\section{Conclusion}\label{sec_conclusion}

This work introduced a quantum framework for image super-resolution based on ANO-VQCs. By jointly optimizing variational parameters and trainable Hermitian observables, the proposed model extends the representational power of conventional VQCs, enabling adaptive multi-qubit measurements that capture fine-grained spatial correlations. 
Experiments on the MNIST dataset demonstrated that ANO--VQCs achieve effective reconstruction of high-resolution images from highly compressed inputs, with quantitative improvements in MSE, PSNR, and SSIM compared to fixed-observable counterparts. 
While 3-local observables enhance structural fidelity, a slight increase in LPIPS indicates a perceptual trade-off, highlighting the tunable balance between sharpness and visual realism.

These findings suggest that adaptive measurement design can serve as a powerful mechanism for resource-efficient quantum learning models. 
Future work will explore scaling ANO-VQCs to larger quantum systems, integrating hybrid classical-quantum postprocessing, and extending this approach to more complex datasets and generative vision tasks.

\bibliographystyle{IEEEbib}
\bibliography{references_ANO_SR,bib/qml_examples,bib/vqc,bib/explain_qml}

@inproceedings{zhang2018unreasonable,
  title     = {The Unreasonable Effectiveness of Deep Features as a Perceptual Metric},
  author    = {Zhang, Richard and Isola, Phillip and Efros, Alexei A. and Shechtman, Eli and Wang, Oliver},
  booktitle = {Proceedings of the IEEE Conference on Computer Vision and Pattern Recognition (CVPR)},
  year      = {2018},
  pages     = {586--595}
}

@article{mcclean2016theory,
  title={The theory of variational hybrid quantum-classical algorithms},
  author={McClean, Jarrod R and Romero, Jonathan and Babbush, Ryan and Aspuru-Guzik, Al{\'a}n},
  journal={New Journal of Physics},
  volume={18},
  number={2},
  pages={023023},
  year={2016},
  publisher={IOP Publishing}
}

@article{wang2021survey,
  title={Deep learning for image super-resolution: A survey},
  author={Wang, Z. and Chen, J. and Hoi, S.C.H.},
  journal={IEEE Transactions on Pattern Analysis and Machine Intelligence},
  volume={43},
  number={10},
  pages={3365--3387},
  year={2021}
}

@inproceedings{dong2015srcnn,
  title={Image super-resolution using deep convolutional networks},
  author={Dong, Chao and Loy, Chen Change and He, Kaiming and Tang, Xiaoou},
  booktitle={IEEE Transactions on Pattern Analysis and Machine Intelligence},
  volume={38},
  number={2},
  pages={295--307},
  year={2015}
}

@inproceedings{dong2016fsrcnn,
  title={Accelerating the super-resolution convolutional neural network},
  author={Dong, Chao and Loy, Chen Change and Tang, Xiaoou},
  booktitle={ECCV},
  pages={391--407},
  year={2016},
  organization={Springer}
}

@inproceedings{kim2016vdsr,
  title={Accurate image super-resolution using very deep convolutional networks},
  author={Kim, Jiwon and Lee, Jung Kwon and Lee, Kyoung Mu},
  booktitle={CVPR},
  pages={1646--1654},
  year={2016}
}

@inproceedings{ledig2017srgan,
  title={Photo-realistic single image super-resolution using a generative adversarial network},
  author={Ledig, Christian and Theis, Lucas and Husz{\'a}r, Ferenc and Caballero, Jose and Cunningham, Andrew and Acosta, Alejandro and Aitken, Andrew and Tejani, Alykhan and Totz, Johannes and Wang, Zehan and Shi, Wenzhe},
  booktitle={CVPR},
  pages={4681--4690},
  year={2017}
}

@inproceedings{zhang2018rcan,
  title={Image super-resolution using very deep residual channel attention networks},
  author={Zhang, Yulun and Li, Kunpeng and Li, Kai and Wang, Lichen and Zhong, Bineng and Fu, Yun},
  booktitle={ECCV},
  pages={286--301},
  year={2018}
}

@inproceedings{tong2017dscn,
  title={Image super-resolution using dense skip connections},
  author={Tong, Tong and Li, Gen and Liu, Xiejie and Gao, Qinquan},
  booktitle={ICCV},
  pages={4799--4807},
  year={2017}
}

@inproceedings{niu2020han,
  title={Single image super-resolution via a holistic attention network},
  author={Niu, Bin and Wen, Wengang and Ren, Wenqi and Zhang, Xiangyu and Yang, Liang and Wang, Shiqi and Zhang, Kai and Cao, Xiaochun and Shen, Heng},
  booktitle={ECCV},
  pages={191--207},
  year={2020}
}

@inproceedings{liang2021swinir,
  title={SwinIR: Image restoration using swin transformer},
  author={Liang, Jingyun and Cao, Jiezhang and Sun, Guolei and Zhang, Kai and Van Gool, Luc and Timofte, Radu},
  booktitle={ICCV},
  pages={1833--1844},
  year={2021}
}

@inproceedings{wang2018esrgan,
  title={ESRGAN: Enhanced super-resolution generative adversarial networks},
  author={Wang, Xintao and Yu, Ke and Wu, Shixiang and Gu, Jinjin and Liu, Yihao and Dong, Chao and Qiao, Yu and Loy, Chen Change},
  booktitle={ECCV Workshops},
  year={2018}
}

@article{saharia2022sr3,
  title={Image super-resolution via iterative refinement},
  author={Saharia, Chitwan and Ho, Jonathan and Chan, William and Salimans, Tim and Fleet, David J. and Norouzi, Mohammad},
  journal={IEEE Transactions on Pattern Analysis and Machine Intelligence},
  volume={45},
  number={4},
  pages={4713--4726},
  year={2022}
}

@article{yang2010image,
  title={Image super-resolution via sparse representation},
  author={Yang, Jianchao and Wright, John and Huang, Thomas and Ma, Yi},
  journal={IEEE Transactions on Image Processing},
  volume={19},
  number={11},
  pages={2861--2873},
  year={2010}
}

@inproceedings{yang2014benchmark,
  title={Single-image super-resolution: A benchmark},
  author={Yang, Chih-Yuan and Ma, Chao and Yang, Ming-Hsuan},
  booktitle={ECCV},
  pages={372--386},
  year={2014}
}

@article{liu2023ediffsr,
  title={EDiffSR: An efficient diffusion probabilistic model for remote sensing image super-resolution},
  author={Liu, Xiao and Yuan, Qiang and Jiang, Kai and He, Jingwen and Jin, Xin and Zhang, Liang},
  journal={IEEE Transactions on Geoscience and Remote Sensing},
  year={2023}
}

@article{chung2022mri,
  title={MR image denoising and super-resolution using regularized reverse diffusion},
  author={Chung, Hyungjin and Lee, Eun Sun and Ye, Jong Chul},
  journal={IEEE Transactions on Medical Imaging},
  volume={42},
  number={4},
  pages={1207--1218},
  year={2022}
}

@article{cerezo2021vqa,
  title={Variational quantum algorithms},
  author={Cerezo, M. and Arrasmith, A. and Babbush, R. and Benjamin, S.C. and Endo, S. and Fujii, K. and McClean, J.R. and Mitarai, K. and Yuan, X. and Cincio, L. and Coles, P.J.},
  journal={Nature Reviews Physics},
  volume={3},
  number={9},
  pages={625--644},
  year={2021}
}

@article{bharti2022nisq,
  title={Noisy intermediate-scale quantum algorithms},
  author={Bharti, Kishor and Cervera-Lierta, Alba and Kyaw, The Htet and Haug, Tobias and Alperin-Lea, Shahn and Anand, Abhinav and Degroote, Matthias and Heimonen, Henri and Kottmann, Jakob S. and Menke, Tim and others},
  journal={Reviews of Modern Physics},
  volume={94},
  number={1},
  pages={015004},
  year={2022}
}

@article{lin2025ano,
  title={Adaptive Non-local Observable on Quantum Neural Networks},
  author={Lin, Hsin-Yi and Tseng, Huan-Hsin and Chen, Samuel Yen-Chi and Yoo, Shinjae},
  journal={arXiv preprint arXiv:2501.05663},
  year={2025}
}

@inproceedings{chen2025learning,
  title={Learning to measure quantum neural networks},
  author={Chen, Samuel Yen-Chi and Tseng, Huan-Hsin and Lin, Hsin-Yi and Yoo, Shinjae},
  booktitle={2025 IEEE International Conference on Acoustics, Speech, and Signal Processing Workshops (ICASSPW)},
  pages={1--5},
  year={2025},
  organization={IEEE}
}

\end{document}